\documentclass[11pt,letterpaper]{article}
\usepackage[utf8]{inputenc}

\usepackage{fullpage}
\usepackage[utf8]{inputenc}
\usepackage{amsfonts}
\usepackage{amsthm}
\usepackage{mathrsfs}
\usepackage{amssymb}
\usepackage{hyperref}
\usepackage{amsmath}
\usepackage{cite}
\usepackage{xcolor}
\usepackage{graphicx}
\usepackage{algorithm}
\usepackage{algorithmicx}
\usepackage{cleveref}
\usepackage{authblk}

\usepackage[noend]{algpseudocode}

\newtheorem{definition}{Definition}

\title{Byzantine Agreement with Less Communication: Recent Advances}
\author{Shir Cohen}
\author{Idit Keidar}
\author{Oded Naor} 
\affil{Technion}

\date{}

\begin{document}

\maketitle
{\let\thefootnote\relax\footnote{{Copyright, \copyright~Shir Cohen, Idit Keidar, Oded Naor, 2021}}}

\begin{abstract}
In recent years, Byzantine Agreement is being considered in increasing scales due to the proliferation of blockchains and other decentralized financial technologies.
Consequently, a number of works have improved its communication complexity in various network models.
In this short paper we survey recent advances and outline some open research questions on the subject.
\end{abstract}

\section{Introduction}
The development of reliable distributed systems often relies on Byzantine Agreement (BA)~\cite{lamport2019byzantine}.
In this problem, a set of correct processes aim to reach a common decision, despite the presence of malicious ones.
BA has been around for four decades, yet practical use-cases for it in large-scale systems have emerged only in the last decade. One major application for BA is cryptocurrencies. For example,
Bitcoin~\cite{nakamoto2012bitcoin}, the first cryptocurrency, requires a large set of users to agree on the state of the blockchain. Since Bitcoin is a real currency with real value, the need to protect it against Byzantine users is crucial.
Following Bitcoin, many other blockchains and FinTech platforms have emerged, e.g.,~\cite{nakamoto2012bitcoin,wood2014ethereum,Algorand,baudet2019state}.
Consequently, an efficient implementation, in terms of communication, has become one of the main foci of BA solutions.

It has been shown by Dolev and Reischuk~\cite{DolevBound} that in deterministic algorithms with $n$ processes, $\Omega(n^2)$ word complexity is needed in the worst-case, assuming the number of faulty processes in the system is $f = O(n)$.
The word complexity of a deterministic BA protocol is defined as the number of words all correct processes send until a decision is reached, where a word is a constant number of bits (e.g., the size of a PKI signature).
Nevertheless, almost all deterministic works incur a word complexity of at least $O(n^3)$ in a synchronous model with $n$ processes, $f$ of which can fail, and the optimal resilience of $n \geq 2f+1$~\cite{dolev1983authenticated, abraham2020sync}. 
In fact, this gap remained open for 35 years until recently Momose and Ren~\cite{momose2020optimal} solved synchronous BA with optimal resilience with $O(n^2)$ words, although less resilient solutions with $O(n^2)$ complexity have been known prior to this result.

Yet synchronous solutions are not robust in large-scale systems, where messages can be delayed for extensive periods.
A more practical approach is to consider the \emph{eventual synchrony} (ES) model, where communication is initially asynchronous but eventually becomes synchronous.
Eventually, synchronous algorithms always ensure safety, but their liveness is conditioned on communication becoming timely.
In this model, performance is measured during the synchronous period, and the optimal resilience is $n=3f+1$.
Recent works have used threshold cryptography in order to achieve quadratic complexity in certain optimistic scenarios~\cite{Abraham2018HotStuffTL,baudet2019state,yin2019hotstuff,guo2020dumbo,momose2020optimal}.

Because attackers may cause communication delays, an even more robust approach is to consider a fully asynchronous model. But in this model, BA cannot be solved deterministically~\cite{fischer1985impossibility}.
Although the complexity lower bound does not apply to randomized algorithms, until fairly recently, randomized solutions also required (expected) $O(n^2)$ word complexity~\cite{rabin1983randomized,cachin2005random,mostefaoui2015signature,abraham2019asymptotically}.

A few recent studies have used randomness to circumvent Dolev and Reischuk's lower bound and provide BA solutions with sub-quadratic word complexity~\cite{king2011breaking,Algorand,nakamoto2012bitcoin,cohen2020not,Abraham2018HotStuffTL,baudet2019state,kwon2014tendermint,naor2019cogsworth,spiegelman2020search} in all three models.
There are two approaches to using probability.
The first is weakening the problem guarantees to probabilistic ones. Works in this vein usually utilize committee sampling, assume an adaptive adversary, and provide probabilistic safety and liveness~\cite{king2011breaking,Algorand,nakamoto2012bitcoin,cohen2020not}.
The second considers models in which deterministic BA solutions are possible, and designs protocols where the \emph{expected} complexity is sub-quadratic\cite{Abraham2018HotStuffTL,baudet2019state,kwon2014tendermint,naor2019cogsworth,spiegelman2020search}. The latter works are resilient only to a static adversary whereas the former tolerate a dynamic one.

Each work in the field examines BA through a slightly different lens: BA can be solved in models varying in their timing assumptions, adversarial behaviors, and properties of their outcomes. This work overviews the latest results on the subject. We point out common patterns in BA algorithms and highlight novel techniques. While doing so, we also identify gaps in existing works and raise open questions for future work. We start by defining the BA problem and describing possible models in Section \ref{sec:model}. In Section \ref{sec:deterministic}, we cover works for deterministic BA, and in Section \ref{sec:randomized} for the randomized problem. We end with some open questions in Section \ref{sec:conclusions}.
\section{Models and Definitions}
\label{sec:model}
We discuss different variations of the \emph{Byzantine Agreement (BA)} problem also known as \emph{consensus}. In this paper, we use the terms consensus and BA interchangeably. In this problem, a set $\Pi$ of $n$ processes attempt to reach a common decision upon one of the values initially suggested by the processes in $\Pi$. Different models affect the problem's difficulty. The connecting thread of our discussions is the presence of up to $f$ Byzantine processes in $\Pi$. These processes may deviate arbitrarily from the protocol, possibly being corrupted and controlled by an \emph{adversary}. Processes in $\Pi$ that are not corrupted are called \emph{correct} processes.
Let us define the problem formally.

\begin{definition}[Byzantine Agreement]
\label{def:BA}
In Byzantine Agreement, each correct process $p_i\in\Pi$ proposes an input value $v_i$ and decides on an output value $decision_i$ s.t. the following properties hold:

\begin{description}
    \item [Validity] If all correct processes propose the same value $v$, then all correct processes decide $v$.
    \item [Agreement] No two correct processes decide differently.
    \item [Termination] Every correct process eventually decides.
\end{description}
\end{definition}

Slight changes to the definition above give rise to several variations of the problem, applicable to different settings and areas of interest. For example, the problem as defined above does not make any assumption on processes' input values, and is known as \emph{multi-value} BA. If we were to require that $v_i$ is a binary value, the problem would be \emph{binary BA}.
The validity condition specified above is known as \emph{strong unanimity}, and it might be more suitable for binary BA (since there are only two possible values).

For multi-value BA, a validity condition named \emph{external validity} was proposed by Cachin et al.~\cite{CachinSecure}.
In this condition, there is some global predicate $\textit{validate}(v) \in \{\textit{true}, \textit{false}\}$ that can be computed locally.
If a correct process decides a value $v$, then $\textit{validate}(v) = \textit{true}$.
The external validity condition is more suitable for multi-value BA, and is used in recent multi-value BA algorithms such as~\cite{yin2019hotstuff,OptimalABA,CachinSecure}.
An example for external validity applicability can be found in various blockchains, where the predicate is verifying that the decided blocks meet the criteria of an eligible block to be included in the blockchain (valid signatures on the transactions, no double-spend, etc.).
We note that the Dolev-Reischuk lower bound~\cite{DolevBound} does not apply to the external validity BA variant, and to the best of our knowledge, no similar lower bound was proven for this case.

\Cref{def:BA} introduces  \emph{deterministic BA}, which requires that the properties hold in all runs of the protocol.
In contrast, in \emph{randomized BA}, one or more of the properties is ensured only with high probability (WHP). In some cases, randomized BA requires deterministic validity and agreement while allowing termination with probability 1. Other works~\cite{Algorand,king2011breaking,cohen2020not} solve \emph{BA with high probability}, where all properties are assured with high probability.

\paragraph{Timing assumptions}
The BA problem(s) can be examined under different timing assumptions. In a \emph{synchronous} network, there is a bound $\delta$ on the time it takes a message to reach its destination. In the \emph{asynchronous} model, we assume nothing about communication times, so message delays are unbounded.
\emph{Eventual synchrony} is a middle ground where the system first behaves asynchronously, and eventually becomes synchronous. The asynchronous period is unbounded but it is guaranteed that at some point in time, called the \emph{Global Stabilization Time (GST)}, the system becomes stable.

\paragraph{Adversary models}
The adversary may corrupt up to $f$ processes in the system. It is known that to solve BA in a synchronous model with strong unanimity, $n$ must be at least $2f+1$.
Under eventual synchrony, the condition is stronger, and $n\geq 3f+1$ is required~\cite{dwork1988consensus}. 

The number of Byzantine processes corrupted by the adversary, known as the \emph{resilience} of the protocol, is not the only concern for the protocol designer. Another important aspect is when and how the adversary controls them. We distinguish between a \emph{static} adversary and an \emph{adaptive} one. The static adversary decides which processes to corrupt before the execution begins, without prior knowledge of processes' initial values or coin flips. An adaptive adversary, on the other hand, can corrupt processes at any time, as long as it does not corrupt more than the threshold $f$.

Getting into a higher resolution, we can also examine the information available to the adversary while it makes its decisions and its power to prevent information flow in the system. In a synchronous model, one considers \emph{“after-the-fact” removals}~\cite{abraham2019communication}. An adversary that is capable of after-the-fact removals can first observe messages sent by processes and then remove these messages from the network by corrupting their senders. It was shown that no after-the-fact removals are required for achieving sub-quadratic word complexity~\cite{abraham2019communication}. The idea of preventing the adversary from “front running” messages sent by processes before corrupting their senders can also be formulated in the asynchronous model.
It can be assumed either explicitly~\cite{cohen2020not} or implicitly by using a separate key to encrypt each message and atomically deleting the key upon sending a message signed by it~\cite{Algorand,blumasynchronous}.
Additionally, one can use the notion of causality to define a \emph{delayed-adaptive} adversary~\cite{cohen2020not}. This adversary can use the contents of a message $m$ sent by a correct process for scheduling a message $m'$ only if $m$ causally precedes $m'$.

\paragraph{Resilience}
The resilience is the upper bound on the number of Byzantine processes in which the problem is still solvable.
In eventual synchrony and asynchronous networks, the optimal resilience is $f < n/3$, independent of the validity conditions or the adversarial model described above.
In the synchrony model, optimal resilience with strong unanimity is $f<n/2$~\cite{momose2020optimal}.
With external validity, the optimal resilience in synchronous networks is $f<n$.

\paragraph{Complexity of Byzantine agreement}
The way we measure complexity depends on the timing assumptions.

\begin{description}
    \item[Synchrony] Under synchrony assumptions, the \emph{worst-case} word complexity is defined as the maximum number of words all the correct processes send under any possible adversarial behavior allowed by the model.
    This is counted from the beginning of the run and until the last correct process reaches a decision.
    
    \item[Eventual synchrony] In the eventual synchrony model, the definition is usually identical to the above, except that the number of words is measured from GST and until a decision is reached.
    Sometimes, the complexity is defined by observing synchronous runs.
    
    \item[Asynchrony] Since asynchronous deterministic consensus is not achievable even in the presence of at least one failure, there is no way to bound the worst-case complexity.
    Instead, we consider the \emph{expected}  complexity.
    The expectation is over the distribution of randomness used to solve BA with probability 1, assuming that the adversary takes the actions that lead to the worst complexity under this distribution.
\end{description}
\begin{figure}[t]
\centering
\includegraphics[width=1.0\textwidth]{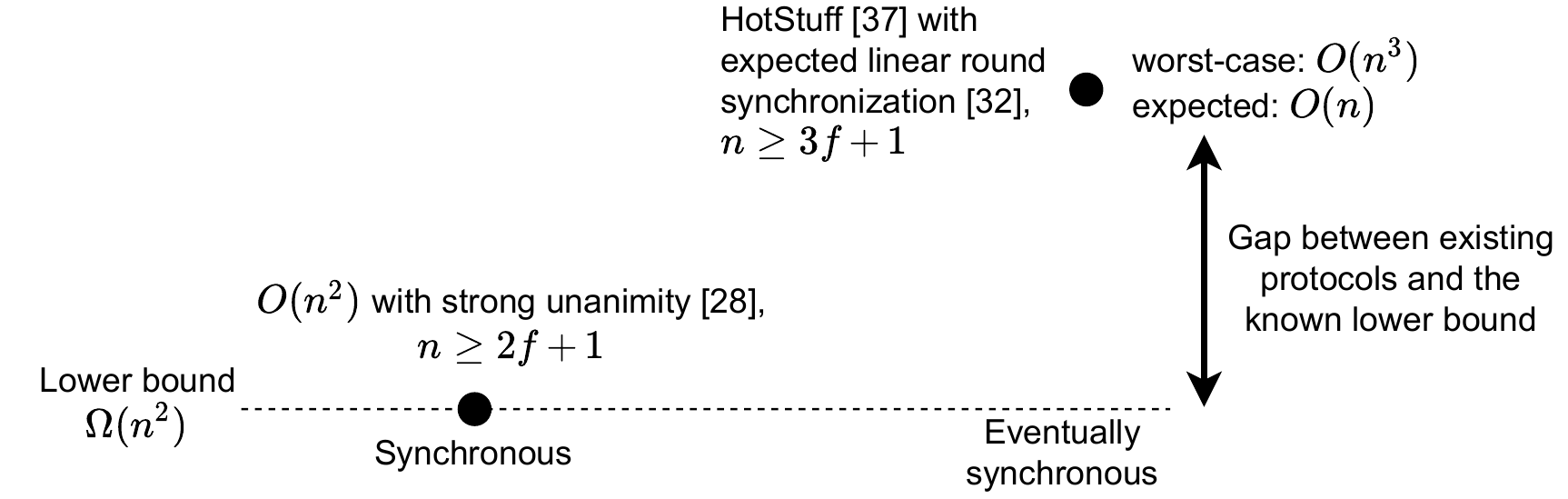}
\caption{Illustration of the gap between existing BA protocols in the eventual synchrony and synchrony models and the known lower bound.}
\label{fig:deterministicGap}
\end{figure}

\section{Deterministic Consensus}
\label{sec:deterministic}
In deterministic consensus protocols, the liveness property, i.e., the property that all processes eventually decide, is guaranteed.
The seminal result of FLP~\cite{fischer1985impossibility} showed that deterministic consensus under asynchrony assumptions in the presence of at least one faulty process cannot be achieved.
Thus, deterministic consensus protocols assume either a completely synchronous network during the entire run, or they assume sufficiently long times in which the network is synchronous, that are used to make progress in the consensus protocol.

In recent years, an increasing number of works such as~\cite{Abraham2018HotStuffTL,baudet2019state,yin2019hotstuff,guo2020dumbo,momose2020optimal} have used threshold cryptography to reduce the word complexity of BA protocols.
This approach works as follows:
A $K$-of-$n$ threshold signature is used to create a signature on some message $m$.
Each of the $n$ processes has a secret key, which it uses to create a unique share of the signature.
Any $K$ shares can be combined into a valid cryptographic signature on $m$.
Each individual share and the combined signature consist of a single word.
In BA protocols, usually all parties send a share of a message $m$ to a process $p$, and once $p$ has $K$ shares (usually $K = f+1$ or $K = 2f+1$), it combines them into a signature and sends it to all processes.
Thus, $p$ can prove that a certain threshold of processes signed $m$ with only $O(n)$ complexity.

Dolev and Reischuk proved in the early 1980s~\cite{DolevBound} that the lower bound for BA in the worst-case is at least $\Omega(n^2)$ messages per decision.
To the best of our knowledge, until recently no deterministic BA protocol in the synchronous model or the eventual synchronous model met this lower bound with optimal resilience.

Recently, Momose and Ren~\cite{momose2020optimal} showed a protocol for BA with strong unanimity with $O(n^2)$ worst-case word complexity in the synchronous model with optimal resilience.
Their protocol is based on threshold cryptography and ideas from~\cite{berman1992bit}, which also solves the problem with $O(n^2)$ communication complexity but for $f<n/3$, i.e., sub-optimal resilience.
While there are synchronous BA protocols with external validity such as~\cite{abraham2020sync}, to the best of our knowledge, there is no known protocol with the same resilience or better and $O(n^2)$ message complexity for solving BA with external validity.

In the eventually synchronous model, recent works such as Tendermint~\cite{kwon2014tendermint}, Casper~\cite{buterin2017casper}, and HotStuff~\cite{yin2019hotstuff} solve BA in $O(n)$ word complexity in optimistic scenarios, sometimes called the ``happy path''.
In the worst-case, HotStuff requires $O(n^3)$ communication (after GST).
Specifically, HotStuff's complexity is $O(n)$ whenever all correct processes are synchronized to the same round and follow a correct leader in this round.
Achieving this is known as the round synchronization problem~\cite{naor2020expectedLinear}.
In HotStuff, the details of the round synchronization are abstracted away into a module named a Pacemaker, whose implementation is left unspecified.
LibraBFT~\cite{baudet2019state}, which is based on HotStuff, implements the Pacemaker with cubic complexity in the worst-case.

A recent work by Naor and Keidar~\cite{naor2020expectedLinear}, which is based on~\cite{naor2019cogsworth}, uses randomization to solve round synchronization with expected $O(n)$ word complexity in the ES setting under a static adversary.
By combining this protocol with HotStuff, multi-shot BA, sometimes referred to as State Machine Replication (SMR), can be solved with expected $O(n)$ in the average case, but with $O(n^3)$ word complexity in the worst-case.
It is still unknown if in the ES model under an adaptive adversary, BA can be solved with $O(n^2)$ word complexity in the worst-case.

The gap between known results and the lower bound for deterministic BA is illustrated in \Cref{fig:deterministicGap}.
\section{Randomized Consensus}
\label{sec:randomized}
In randomized BA algorithms, the execution depends on local coin tosses at processes. While a simple local coin is a basic building block used in traditional works~\cite{ben1983another,bracha1983resilient}, simply tossing coins may result in an exponential expected time to reach agreement.
A more efficient approach is to utilize some sort of local randomness in order to create a \emph{shared coin}. Processes communicate with each other to try to agree on the coin toss, and the coin has some well-defined \emph{success rate}~\cite{rabin1983randomized,canetti1993fast,cachin2005random,Algorand,king2013byzantine}, namely a positive probability for all to agree on a random bit. 

In previous sections, we mentioned Dolev and Reischuk’s renowned lower bound, which indicates the necessity of quadratic communication \cite{DolevBound} in deterministic settings. Can we do better by using randomized algorithms? The answer depends on the adversary. First, if we assume an adaptive adversary  that is capable of after-the-fact removal, the answer is no. Abraham et al.~\cite{abraham2019communication} recently extended the classic bound and proved that any randomized BA protocol under such an adversary must incur quadratic communication in expectation. Furthermore, a number of works~\cite{abraham2019communication,blumasynchronous,Rambaud} have proved that for subquadratic BA some form of setup assumption is required.

On the other hand, some algorithms that assume a restricted adversary and trusted setup do break the barrier of quadratic communication, solving binary BA. A common way of reducing algorithms' communication complexity is using committee sampling. With this technique, each step of the protocol is executed by a subset of the processes. That is, a process that is sampled to a committee related to a specific message sends it, whereas a process that is not sampled does nothing. In the presence of an adaptive adversary, we must ensure that the sampling for each committee cannot be anticipated by the adversary. To this end, committees are independent of each other, and messages are sent to all processes.

The first sub-quadratic algorithm was introduced by King and Saia~\cite{king2011breaking} in the full information synchronous model, where there are no restrictions on the adversary’s computational power. While they manage to achieve $\widetilde{O}(n\sqrt{n})$ word complexity, their work achieves very low resilience ($n>400f$).

By using cryptographic abstractions and assuming a computationally bounded adversary, it is possible to achieve both high resilience and low communication complexity. This has been done by using \emph{verifiable random functions} (VRFs)~\cite{micali1999VRF} to elect committees via a procedure known as \emph{cryptographic sortition}~\cite{Algorand}.
A VRF is a pseudorandom function that provides a proof of its correct computation.
Based on a local VRF computation, processes can
choose committee members without communicating with each other. Each process runs a local procedure, independent of other processes, which checks whether it is elected. If selected, it is provided with a proof that can be used as a certificate to others. The unpredictability of the selection is a protective measure against an adaptive adversary that seeks to corrupt the elected members.

An adversary can identify correct processes that participate in a committee once they send messages. An unrestricted adaptive adversary can then corrupt a committee member and schedule a new message to arrive before the message which was sent before the corruption. We recall that this scenario can be prevented by an explicit assumption~\cite{cohen2020not}, or implicitly in the so-called \emph{erasure model}~\cite{Algorand,blumasynchronous}, where correct processes are able to securely erase secret keys from memory.
This matches the no after-the-fact lower bound in~\cite{abraham2019communication}, as the adversary cannot conceal messages sent by processes that were not corrupted at the time of the sending.

Algorand~\cite{Algorand} first defined and used cryptographic sortition.
It solves BA WHP with subquadratic communication and a near-optimal resilience of $f=(\frac{1}{3}-\epsilon)n$, assuming a trusted \emph{public key infrastructure} (PKI) in the erasure model and using VRFs to sample committees. Their initial work was presented in a synchronous model, and was followed by a solution assuming eventual synchrony~\cite{chen2018algorand}. In both cases, they rely on the timing assumption to make progress.
Specifically, their shared coin implementation is based on the idea that every value sent by a correct process reaches all processes in the system before a timeout signals the end of a ``round''. This is clearly not the case in asynchronous settings.

Algorand's shared coin implementation is quite simple and based on VRF values. A simplified pseudo-code of it appears in Algorithm~\ref{alg:algorand}. In their coin, every process broadcasts a pseudo-random value originated at a VRF output.
At the end of the communication round, the minimum value is evaluated. The chosen random bit is the minimum value's least significant bit. Whenever the minimum value is sent by a correct process, all processes receive it and result in the same coin toss. Since the VRFs are pseudo-random and verified, there is a probability of at least $\frac{2}{3}$ that a correct process holds the minimum value and all correct processes obtain the same bit.

\begin{algorithm}
    \begin{algorithmic}[1]
    \State $v_{min}\gets \infty$
    \State
    $v_i \gets \emph{VRF}_i(r)$
    \State send ${v_i}$ to all processes
    
    \State \textbf{wait} until timeout
    
    \For{all $v_j$ received in round $r$}
    \State $v_{min}\gets \min\{v_{min},v_j\}$
    \EndFor
    \State return $LSB(v_{min})$

    \end{algorithmic}
    \caption{Algorand's Shared Coin Framework - coin toss in round $r$}
    \label{alg:algorand}
\end{algorithm}

Recently, the quadratic barrier was broken also for the asynchronous case. Cohen et al.~\cite{cohen2020not} have solved BA WHP in a model that assumes a trusted PKI and a delayed-adaptive adversary that cannot “front run” messages sent by processes before corrupting their sender. A simplified pseudo-code of their shared coin appears in Algorithm~\ref{alg:coincidence}.
With no timing assumptions, the task of handling dynamically selected committees is more difficult and the notion of ``timeout'' loses its meaning. Instead, they condition the protocol's progress on a predefined parameter $W$.
They present sufficient conditions on sampling, which ensure safety and liveness with high probability. Specifically, the conditions ensure that waiting for $W$ messages allows progress WHP.
Because the adversary can create disagreement by hiding messages of correct processes from each other, achieving a positive success rate is more challenging. It requires an additional communication round for ``boosting'' the presence of the values. In addition, their shared coin algorithm restricts the protocol's resilience to $n\approx4.5f$.
We note that the delayed adaptiveness of the adversary is necessary to ensure that it cannot discover who holds the global minimum before scheduling the messages. Without it, the adversary would be able to hide the minimum from some correct processes, creating disagreement.

\begin{algorithm}
    \begin{algorithmic}[1]
    \State $v_{min}\gets \infty$
    \State
    $v_i \gets \emph{VRF}_i(r)$
    \State send ${v_i}$ to all processes
    
    \State \textbf{wait} until receiving \textbf{$W$} messages
    
    \For{all $v_j$ received in the $W$ messages}
    \State $v_{min}\gets \min\{v_{min},v_j\}$
    \EndFor
    \State send ${v_{min}}$ to all processes

    \State \textbf{wait} until receiving \textbf{$W$} messages
    
    \For{all $v_j$ received in the $W$ messages}
    \State $v_{min}\gets \min\{v_{min},v_j\}$
    \EndFor
    \State return $LSB(v_{min})$

    \end{algorithmic}
    \caption{Cohen et al.'s Shared Coin Framework - coin toss in round $r$}
    \label{alg:coincidence}
\end{algorithm}

Blum et al.~\cite{blumasynchronous} have presented an algorithm that solves asynchronous BA with near-optimal resilience. Compared to Cohen et al., they strengthen the adversary, removing the delayed adaptiveness requirement, but also the trusted setup phase. Their trusted dealer fixes committees' members and coin tosses in advance. Furthermore, it distributes the tosses to the members before the execution starts using secret sharing. As in Algorand, they also work in the erasure model. Algorithm~\ref{alg:blum} presents a simplified pseudo-code of their shared coin.

\begin{algorithm}
    \begin{algorithmic}[1]
    \State send predefined round $r$ share to all processes
    \State return constructed predefined round $r$ secret

    \end{algorithmic}
    \caption{Blum et al.'s Shared Coin Framework - coin toss in round $r$}
    \label{alg:blum}
\end{algorithm}

Another technique that can used to lower the word complexity of SMR algorithms (but not single-shot BA) is batching of values, i.e., a process decides on multiple values in each decision event.
By doing so, the \emph{amortized} word complexity of each single decision can be lowered. 

HoeneyBadgerBFT~\cite{miller2016honey} is an asynchronous SMR algorithm with amortized linear word complexity per decision, that relies on cryptographic threshold signatures for safety.
DAG-Rider~\cite{keidar2021allDAG}, a recent asynchronous multi-value SMR algorithm achieves amortized linear word complexity, and does not rely on any cryptographic assumptions for its safety.
DAG-Rider represents network communication using a Directed Acyclic Graph (DAG), where each vertex is a message broadcast by some process, containing a batch of proposed consensus values. 
The DAG vertices are reliably broadcast.
On top of the DAG, there is a local ordering layer that uses a perfect shared coin to decide on a sequence of vertices, in such a way that all correct processes agree 
on the same vertices in the same order. By batching proposals in each vertex, each single decision has only an expected linear word complexity.
\section{Open Questions}
\label{sec:conclusions}

While in recent years there are significant improvements in the word complexity of Byzantine agreement, we identify some still unanswered questions:

\begin{itemize}
   \item Is there a lower bound on the number of messages or signatures for BA with external validity, similar to Dolev-Reischuk's~\cite{DolevBound} $\Omega(n^2)$ lower bound?
    
    \item In the synchronous model, is there a multi-value BA protocol with external validity that has $O(n^2)$ word complexity in the worst-case?
        
    \item In the eventual synchrony setting, is there any BA protocol (binary or multi-value) that has $O(n^2)$ word complexity in the worst-case?
    
    \item Can sub-quadratic BA be solved with deterministic guarantees for some of the properties, specifically safety ones, while others hold WHP? 
    
    \item In the asynchronous settings, is there a sub-quadratic multi-value BA protocol?
    
    \item Whilst existing asynchronous sub-quadratic BA protocols rely on guarantees that hold whp provided that the protocol is run a constant number of times, is it possible to design asynchronous sub-quadratic SMR? 
\end{itemize}

\section*{Acknowledgements}
Oded Naor is grateful to the Azrieli Foundation for the
award of an Azrieli Fellowship.

\bibliographystyle{plain}
\bibliography{references}

\begin{thebibliography}{10}

\bibitem{abraham2019communication}
Ittai Abraham, TH~Hubert Chan, Danny Dolev, Kartik Nayak, Rafael Pass, Ling
  Ren, and Elaine Shi.
\newblock Communication complexity of byzantine agreement, revisited.
\newblock In {\em Proceedings of the 2019 ACM Symposium on Principles of
  Distributed Computing}, pages 317--326, 2019.

\bibitem{Abraham2018HotStuffTL}
Ittai Abraham, Guy Golan-Gueta, and Dahlia Malkhi.
\newblock {Hot-Stuff the Linear, Optimal-Resilience, One-Message BFT Devil}.
\newblock {\em CoRR}, abs/1803.05069, 2018.

\bibitem{abraham2020sync}
Ittai Abraham, Dahlia Malkhi, Kartik Nayak, Ling Ren, and Maofan Yin.
\newblock {Sync HotStuff: Simple and practical synchronous state machine
  replication}.
\newblock In {\em 2020 IEEE Symposium on Security and Privacy (SP)}, pages
  106--118. IEEE, 2020.

\bibitem{OptimalABA}
Ittai Abraham, Dahlia Malkhi, and Alexander Spiegelman.
\newblock Validated asynchronous byzantine agreement with optimal resilience
  and asymptotically optimal time and word communication.
\newblock {\em CoRR}, abs/1811.01332, 2018.

\bibitem{abraham2019asymptotically}
Ittai Abraham, Dahlia Malkhi, and Alexander Spiegelman.
\newblock Asymptotically optimal validated asynchronous byzantine agreement.
\newblock In {\em Proceedings of the 2019 ACM Symposium on Principles of
  Distributed Computing}, pages 337--346, 2019.

\bibitem{baudet2019state}
Mathieu Baudet, Avery Ching, Andrey Chursin, George Danezis, Fran{\c{c}}ois
  Garillot, Zekun Li, Dahlia Malkhi, Oded Naor, Dmitri Perelman, and Alberto
  Sonnino.
\newblock {State machine replication in the Libra Blockchain}.
\newblock {\em The Libra Assn., Tech. Rep}, 2019.

\bibitem{ben1983another}
Michael Ben-Or.
\newblock Another advantage of free choice (extended abstract): Completely
  asynchronous agreement protocols.
\newblock In {\em Proceedings of the second annual ACM symposium on Principles
  of distributed computing}, pages 27--30. ACM, 1983.

\bibitem{berman1992bit}
Piotr Berman, Juan~A Garay, and Kenneth~J Perry.
\newblock Bit optimal distributed consensus.
\newblock In {\em Computer science}, pages 313--321. Springer, 1992.

\bibitem{blumasynchronous}
Erica Blum, Jonathan Katz, Chen-Da Liu-Zhang, and Julian Loss.
\newblock Asynchronous byzantine agreement with subquadratic communication.
\newblock Cryptology ePrint Archive, Report 2020/851, 2020.
\newblock \url{https://eprint.iacr.org/2020/851}.

\bibitem{bracha1983resilient}
Gabriel Bracha and Sam Toueg.
\newblock Resilient consensus protocols.
\newblock In {\em Proceedings of the second annual ACM symposium on Principles
  of distributed computing}, pages 12--26. ACM, 1983.

\bibitem{buterin2017casper}
Vitalik Buterin and Virgil Griffith.
\newblock Casper the friendly finality gadget.
\newblock {\em arXiv preprint arXiv:1710.09437}, 2017.

\bibitem{CachinSecure}
Christian Cachin, Klaus Kursawe, Frank Petzold, and Victor Shoup.
\newblock Secure and efficient asynchronous broadcast protocols.
\newblock In {\em Advances in Cryptology ---CRYPTO 2001}, pages 524--541.
  Springer Berlin Heidelberg, 2001.

\bibitem{cachin2005random}
Christian Cachin, Klaus Kursawe, and Victor Shoup.
\newblock Random oracles in {C}onstantinople: Practical asynchronous byzantine
  agreement using cryptography.
\newblock {\em Journal of Cryptology}, 18(3):219--246, 2005.

\bibitem{canetti1993fast}
Ran Canetti and Tal Rabin.
\newblock Fast asynchronous byzantine agreement with optimal resilience.
\newblock In {\em STOC}, volume~93, pages 42--51. Citeseer, 1993.

\bibitem{chen2018algorand}
Jing Chen, Sergey Gorbunov, Silvio Micali, and Georgios Vlachos.
\newblock Algorand agreement: Super fast and partition resilient byzantine
  agreement.
\newblock {\em IACR Cryptology ePrint Archive}, 2018:377, 2018.

\bibitem{cohen2020not}
Shir Cohen, Idit Keidar, and Alexander Spiegelman.
\newblock Not a coincidence: Sub-quadratic asynchronous byzantine agreement
  whp.
\newblock {\em arXiv preprint arXiv:2002.06545}, 2020.

\bibitem{DolevBound}
Danny Dolev and R\"{u}diger Reischuk.
\newblock Bounds on information exchange for byzantine agreement.
\newblock {\em J. ACM}, 32(1):191–204, January 1985.

\bibitem{dolev1983authenticated}
Danny Dolev and H.~Raymond Strong.
\newblock Authenticated algorithms for byzantine agreement.
\newblock {\em SIAM Journal on Computing}, 12(4):656--666, 1983.

\bibitem{dwork1988consensus}
Cynthia Dwork, Nancy Lynch, and Larry Stockmeyer.
\newblock Consensus in the presence of partial synchrony.
\newblock {\em Journal of the ACM (JACM)}, 35(2):288--323, 1988.

\bibitem{fischer1985impossibility}
Michael~J Fischer, Nancy~A Lynch, and Michael~S Paterson.
\newblock Impossibility of distributed consensus with one faulty process.
\newblock {\em Journal of the ACM (JACM)}, 32(2):374--382, 1985.

\bibitem{Algorand}
Yossi Gilad, Rotem Hemo, Silvio Micali, Georgios Vlachos, and Nickolai
  Zeldovich.
\newblock Algorand: Scaling byzantine agreements for cryptocurrencies.
\newblock In {\em Proceedings of the 26th Symposium on Operating Systems
  Principles}, pages 51--68, 2017.

\bibitem{guo2020dumbo}
Bingyong Guo, Zhenliang Lu, Qiang Tang, Jing Xu, and Zhenfeng Zhang.
\newblock Dumbo: Faster asynchronous bft protocols.
\newblock In {\em Proceedings of the 2020 ACM SIGSAC Conference on Computer and
  Communications Security}, pages 803--818, 2020.

\bibitem{keidar2021allDAG}
Idit Keidar, Eleftherios Kokoris-Kogias, Oded Naor, and Alexander Spiegelman.
\newblock All you need is dag.
\newblock In {\em the future proceedings of the 2021 ACM Symposium on
  Principles of Distributed Computing}, 2021.

\bibitem{king2011breaking}
Valerie King and Jared Saia.
\newblock Breaking the {$O(n^2)$} bit barrier: scalable byzantine agreement
  with an adaptive adversary.
\newblock {\em Journal of the ACM (JACM)}, 58(4):1--24, 2011.

\bibitem{king2013byzantine}
Valerie King and Jared Saia.
\newblock Byzantine agreement in polynomial expected time.
\newblock In {\em Proceedings of the forty-fifth annual ACM symposium on Theory
  of computing}, pages 401--410. ACM, 2013.

\bibitem{kwon2014tendermint}
Jae Kwon.
\newblock Tendermint: Consensus without mining.
\newblock {\em Draft v. 0.6, fall}, 1(11), 2014.

\bibitem{lamport2019byzantine}
Leslie Lamport, Robert Shostak, and Marshall Pease.
\newblock The byzantine generals problem.
\newblock {\em ACM Trans. Program. Lang. Syst.}, 4(3):382–401, July 1982.

\bibitem{micali1999VRF}
Silvio Micali, Michael Rabin, and Salil Vadhan.
\newblock Verifiable random functions.
\newblock In {\em Foundations of Computer Science, 1999. 40th Annual Symposium
  on}, pages 120--130. IEEE, 1999.

\bibitem{miller2016honey}
Andrew Miller, Yu~Xia, Kyle Croman, Elaine Shi, and Dawn Song.
\newblock The honey badger of bft protocols.
\newblock In {\em Proceedings of the 2016 ACM SIGSAC Conference on Computer and
  Communications Security}, pages 31--42, 2016.

\bibitem{momose2020optimal}
Atsuki Momose and Ling Ren.
\newblock Optimal communication complexity of byzantine consensus under honest
  majority.
\newblock {\em arXiv preprint arXiv:2007.13175}, 2020.

\bibitem{mostefaoui2015signature}
Achour Most{\'e}faoui, Hamouma Moumen, and Michel Raynal.
\newblock Signature-free asynchronous binary byzantine consensus with $t <
  n/3$, ${O}(n^2)$ messages, and ${O}(1)$ expected time.
\newblock {\em Journal of the ACM (JACM)}, 62(4):31, 2015.

\bibitem{nakamoto2012bitcoin}
Satoshi Nakamoto.
\newblock Bitcoin: A peer-to-peer electronic cash system, 2009.

\bibitem{naor2019cogsworth}
Oded Naor, Mathieu Baudet, Dahlia Malkhi, and Alexander Spiegelman.
\newblock Cogsworth: Byzantine view synchronization.
\newblock In {\em Proceedings of the Cryptoeconomic Systems Conference
  (CES'20)}, 2020.

\bibitem{naor2020expectedLinear}
Oded Naor and Idit Keidar.
\newblock {Expected Linear Round Synchronization: The missing link for Linear
  Byzantine SMR}.
\newblock In {\em 34th International Symposium on Distributed Computing
  (DISC)}, 2020.

\bibitem{rabin1983randomized}
Michael~O Rabin.
\newblock Randomized byzantine generals.
\newblock In {\em 24th Annual Symposium on Foundations of Computer Science
  (sfcs 1983)}, pages 403--409. IEEE, 1983.

\bibitem{Rambaud}
Matthieu Rambaud.
\newblock Lower bounds for authenticated randomized byzantine consensus under
  (partial) synchrony: The limits of standalone digital signatures.

\bibitem{spiegelman2020search}
Alexander Spiegelman.
\newblock In search for a linear byzantine agreement.
\newblock {\em arXiv preprint arXiv:2002.06993}, 2020.

\bibitem{wood2014ethereum}
Gavin Wood et~al.
\newblock Ethereum: A secure decentralised generalised transaction ledger.
\newblock {\em Ethereum project yellow paper}, 151(2014):1--32, 2014.

\bibitem{yin2019hotstuff}
Maofan Yin, Dahlia Malkhi, Michael~K Reiter, Guy~Golan Gueta, and Ittai
  Abraham.
\newblock {HotStuff: BFT consensus with linearity and responsiveness}.
\newblock In {\em Proceedings of the 2019 ACM Symposium on Principles of
  Distributed Computing}, pages 347--356, 2019.

\end{thebibliography}
\end{document}